\documentclass[aps,prd,groupedaddress,showpacs,floatfix,12pt]{revtex4}
\usepackage{graphicx}
\usepackage{dcolumn}
\usepackage{bm}
\begin{document}
\title{Once more about gauge invariance in \boldmath{$\phi\to\gamma\pi^0\pi^0$}.}
\author{N.~N.~Achasov}
\email{achasov@math.nsc.ru}
\affiliation{Laboratory of  Theoretical Physics, S.~L.~Sobolev Institute for Mathematics,
 Novosibirsk, 630090, Russia}

\date{\today}

\begin{abstract}
It is argued that the realization of gauge invariance condition as
a consequent of cancellation between the $\phi\to\gamma
f_0\to\gamma\pi^0\pi^0 $ resonance contribution
 and a $\phi\to\gamma\pi^0\pi^0$ background one, suggested in Ref. \cite{anisovich},  is
 misleading.
\end{abstract}
\vspace*{1cm}
 \pacs{12.39.-x, 13.40.Hq, 13.65.+i}
\maketitle \vspace*{1cm} Gauge invariance requires that the
$\phi\to\gamma\pi^0\pi^0$ amplitude form be
\begin{equation}
\label{gi}
 A\left (\phi(p)\to\gamma(k)\pi^0(q_1)\pi^0(q_2)\right )= F_{\mu\nu}T^{\mu\nu},
\end{equation}
where $F_{\mu\nu}=k_\mu e_\nu(\gamma) - k_\nu e_\mu(\gamma)$,
$e(\gamma)$ is the $\gamma$ quantum polarization four-vector, the
tensor $T^{\mu\nu}$ includes the $\phi$ meson polarization
four-vector $e(\phi)$.

Since there are no charge particles or particles with magnetic
moments in the process, there is the low energy theorem
\begin{equation}
\label{soft}
 A\left (\phi(p)\to\gamma(k)\pi^0(q_1)\pi^0(q_2)\right
 )\to O(k)\ \mbox{(at least!)},
\end{equation}
when $k\to 0$, ($p=k+q_1+q_2$).

If you calculate the $\phi(p)\to\gamma (k) f_0(q)$ transition
amplitude  in field theory, using gauge invariant method, you get
\cite{achasov}
\begin{eqnarray}
\label{giv}
 && A\left (\phi(p)\to\gamma (k) f_0(q)\right )\sim G_{f_0}F_{\mu\nu}p^\mu e^\nu(\phi)\nonumber\\
 &&= G_{f_0}\left ((p\cdot k)\left (e(\phi)\cdot e(\gamma)\right )-
 \left (k\cdot e(\phi)\right )\left (p\cdot e(\gamma)\right )\right ),
 \end{eqnarray}
 where  $G_{f_0}$ is an invariant vertex free from
kinematical singularities  \cite{virtual}.

When $k\to 0$, ($p=k+q$), there is  the low energy theorem again
\begin{equation}
\label{soft1}
 A\left (\phi(p)\to\gamma(k)f_0(q)\right
 )\to O(k)\ \mbox{(at least!)},
\end{equation}
because  there are also no charge particles or particles with
magnetic moments in the process.

The authors of Ref. \cite{anisovich} constructed  such a model in
which Eq. (\ref{soft1}) does not take place, which is why they
suggested to realize the low energy theorem (\ref{soft}) as a
consequent of cancellation between the $\phi\to\gamma
f_0\to\gamma\pi^0\pi^0 $ resonance contribution
 and a $\phi\to\gamma\pi^0\pi^0$ background one \cite{anisovich}:
 \begin{equation}
 \label{anis}
A\left (\phi(p)\to\gamma(k)\pi^0(q_1)\pi^0(q_2)\right
 )\sim \frac{g_\pi}{M_0^2 - M_{\pi\pi}^2- \imath g_\pi^2\rho_{\pi\pi} - \imath g_K^2\rho_{K\bar
 K}}+ B(M^2_{\pi\pi}),
 \end{equation}
where
\begin{eqnarray}
\label{anisnotification} &&
\rho_{\pi\pi}=\frac{1}{M_0}\sqrt{M_{\pi\pi}^2-4m_\pi^2}\,,\ \
\rho_{K\bar K}=\frac{1}{M_0}\sqrt{M_{\pi\pi}^2-4m_K^2}\,\ \
\mbox{when}\ \ M_{\pi\pi}>2m_K\,,\nonumber\\[6pt]
 && \ \rho_{K\bar
K}=\imath \frac{1}{M_0}\sqrt{4m_K^2-M_{\pi\pi}^2}\,\ \
\mbox{when}\ \ M_{\pi\pi}<2m_K\,, \nonumber\\[6pt]
 &&
g_\pi^2=0.12\,\mbox{GeV}^2,\ \ \ g_K^2=0.27\,\mbox{GeV}^2,\ \ \
M_0=0.975\,\mbox{GeV}.
\end{eqnarray}

The background term is parameterized in the form:
\begin{eqnarray}
\label{background} && B\left (M_{\pi\pi}^2\right )=C\left [ 1 +
a\left (M_{\pi\pi}^2-m_\phi^2\right ) \right]\exp\left [-
\frac{m_\phi^2-m_{\pi\pi}^2}{\mu^2}\right ]\,,\\[6pt]
&&
\mbox{where}\ \ 1/a = - 0.2 \mbox{GeV}^2\,\ \ \mu = 0.388\,
\mbox{GeV}\,.\nonumber
\end{eqnarray}

Note that all notifications and figures in the right side of Eq.
(\ref{anis}) and in Eqs. (\ref{anisnotification}), (
\ref{background}) are taken from Ref. \cite{anisovich}, see Eqs.
(60)-(64) and (66) in Ref. \cite{anisovich} .

The low energy theorem (\ref{soft}) is got by the constraint
\begin{equation}
\label{anisgauge} \left [\, \frac{g_\pi}{M_0^2 - M_{\pi\pi}^2-
\imath g_\pi^2\rho_{\pi\pi} - \imath g_K^2\rho_{K\bar
 K}}+ C\,\right ]_{M_{\pi\pi}=m_\phi}=0\,,
\end{equation}
which leads to the $(photon\ energy)^3=\omega^3$ law for the soft
photon spectrum,  $\omega = ( m_\phi^2 - m_{\pi\pi}^2 )/2m_\phi$.

But the authors of Ref. \cite{anisovich} missed that the low
energy theorem  takes place at every $p^2$, not only for
$p^2=m_\phi^2$, see Eqs. (\ref{gi}), (\ref{soft}). As a result the
model of Ref. \cite{anisovich} pretends to gauge invariance in the
$e+e^-\to\phi\to\gamma\pi^0\pi^0$ reaction only when the total
energy of beams in the center-of-mass system $E=m_\phi$. Really,
\begin{eqnarray}
\label{cross} && \sigma\left ( e+e^-\to\phi\to\gamma\pi^0\pi^0,\
E\right )\sim\Biggl|\,\frac{1}{m_\phi^2 - E^2 - \imath
E\Gamma_\phi(E)}\,\Biggr|^{\,2}\nonumber\\[9pt]
&&\times\Biggl|\,\frac{g_\pi}{M_0^2 - M_{\pi\pi}^2- \imath
g_\pi^2\rho_{\pi\pi} - \imath g_K^2\rho_{K\bar
 K}}+ B\left(M^2_{\pi\pi}\right)\,\Biggr|^{\,2}
\end{eqnarray}
and hence the soft energy theorem $\omega^3$ for the photon
spectrum takes place only at $E=m_\phi$, but not at every $E$ as
gauge invariance requires.

If one tries to use the idea of Ref. \cite{anisovich} about
cancellation between  resonance
 and  background contributions at every $E$, replacing $m_\phi$ in Eq. (\ref{background}) by $E$, one will get the
 discouraging result
 \begin{equation}
\label{everyE}
 C = - \frac{g_\pi}{M_0^2 - E^2 -
\imath g_\pi^2\rho_{\pi\pi}(E) - \imath g_K^2\rho_{K\bar
 K}(E)}
\end{equation}
instead of Eq. (\ref{anisgauge}), where
\begin{eqnarray}
\label{Enotification} &&
\rho_{\pi\pi}(E)=\frac{1}{M_0}\sqrt{E^2-4m_\pi^2}\,,\ \
\rho_{K\bar K}=\frac{1}{M_0}\sqrt{E^2-4m_K^2}\,\ \ \mbox{when}\ \
E>2m_K\,,\nonumber\\[6pt]
 && \ \rho_{K\bar
K}=\imath \frac{1}{M_0}\sqrt{4m_K^2-E^2}\,\ \ \mbox{when}\ \
E<2m_K\,.
\end{eqnarray}
So, a new vector resonance with the mass $E=m_V\approx 980$ MeV
and the visible width $\Gamma_{visible}\approx  70$ MeV  should be
produced together with the $\phi$ meson in the $e+e^-\to V(980) +
\phi\to\gamma\pi^0\pi^0$ reaction:
\begin{eqnarray}
\label{crossv} && \sigma\left ( e+e^-\to
V(980)+\phi\to\gamma\pi^0\pi^0,\ E\right
)\sim\Biggl|\,\frac{1}{m_\phi^2 - E^2 - \imath
E\Gamma_\phi(E)}\nonumber\\[9pt] && \times\left (\frac{1}{M_0^2 -
M_{\pi\pi}^2- \imath g_\pi^2\rho_{\pi\pi} - \imath
g_K^2\rho_{K\bar K}}+ \frac{b}{\Delta}\right )\nonumber\\[9pt]
&&
-\,\frac{1}{M_0^2 - E^2 - \imath g_\pi^2\rho_{\pi\pi}(E) - \imath
g_K^2\rho_{K\bar
 K}(E)}\frac{b}{\Delta}\,\Biggr|^{\,2},
 \end{eqnarray}
where
\begin{eqnarray}
\label{where}
 &&\Delta=m_\phi^2-M_0^2 - \imath E\Gamma_\phi(E)+\imath
g_\pi^2\rho_{\pi\pi}(E) + \imath g_K^2\rho_{K\bar
K}(E)\,,\nonumber\\[6pt] && b=\left [ 1 + a\left
(M_{\pi\pi}^2-E^2\right ) \right]\exp\left [-
\frac{E^2-m_{\pi\pi}^2}{\mu^2}\right ].
\end{eqnarray}

The implication of this story is that every reasonable model
should give the $\phi(p)\to\gamma (k) f_0(q)$ amplitude in the
form of Eq. (\ref{giv}) with a regular $G_{f_0}$ that leads to Eq.
(\ref{soft1}).

Not pretending to a completeness of quoting, let us list a few
papers  about gauge invariance in particle physics \cite{low}
including ones straight related to the process under consideration
\cite{creutz}.

 {\bf
Acknowledgment} This work was supported in part by the
Presidential Grant NSh-5362.2006.2 for Leading Scientific Schools.


\begin{thebibliography}{99}
\bibitem{virtual}
 For a virtual photon, another independent  gauge invariant amplitude arises
\begin{eqnarray}
 \label{virt}
 && A\left (\phi(p)\to\gamma^\ast (k) f_0(q)\right )\sim G_{f_0}F_{\mu\nu}p^\mu e^\nu(\phi)+V_{f_0}F_{\mu\nu}k^\mu e^\nu(\phi)\nonumber\\
  && = G_{f_0}F_{\mu\nu}p^\mu
  e^\nu(\phi)+V_{f_0}\left (k^2\left (e(\gamma)\cdot e(\phi)\right )-\left (k\cdot e(\gamma)\right )\left (k\cdot e(\phi)\right )\right ),
 \end{eqnarray}
 where $V_{f_0}$ is another invariant vertex free from
kinematical singularities.
\end{thebibliography}
\end{document}